\documentclass[aps,prb,showpacs,twocolumn,amsmath,amssymb,superscriptaddress]{revtex4}

\usepackage[colorlinks,citecolor=blue,linkcolor=cyan]{hyperref}
\usepackage[usenames,dvipsnames,svgnames,table]{xcolor}

\usepackage{graphicx}
\usepackage{Jonasmacros}
\usepackage{bm}
\usepackage{mathptmx}
\usepackage{pxfonts}

\begin{document}
\date{\today}
\title{Spin Inelastic Electron Tunneling Spectroscopy on Local Magnetic Moment Embedded in Josephson Junction}

\author{P. Berggren}
\affiliation{Department of Physics and Astronomy, Uppsala University, Box 516, SE-75120, Uppsala, Sweden}

\author{J. Fransson}
\email{jonas.fransson@physics.uu.se}
\affiliation{Department of Physics and Astronomy, Uppsala University, Box 516, SE-75120, Uppsala, Sweden}

\begin{abstract}
Recent experimental conductance measurements performed on paramagnetic molecular adsorbates on a superconducting surface, using superconducting scanning tunneling microscopy techniques, are theoretically investigated. For low temperatures, we demonstrate that tunneling current assisted excitations of the local magnetic moment cannot occur for voltage biases smaller than the superconducting gap of the scanning tunneling microscope. The magnetic moment is only excited for voltages corresponding to the sum of the superconducting gap and the spin excitation energies. In excellent agreement with experiment, we show that pumping into higher excitations give additional current signatures by accumulation of density in the lower ones. Using external magnetic fields, we Zeeman split possible degeneracy and thereby resolve all excitations comprised in the magnetic moment.
\end{abstract}
\pacs{74.55.+v,73.20.Hb,71.70.Gm}

\maketitle

Detecting and protecting information stored in single spin structures have become one of the latest routes to envisage quantum computation. A major challenge is to find suitable systems where the spin excitations have life times long enough to sustain qubit operations. Single spins in contact with a metal surface have short life times \cite{heinrich2004,balashov2009,khajetoorians2011}, in the order of ps or less, due to exchange of energy and angular momentum with the itinerant surface electrons. While this problem was partly overcome by introducing a separating layer, e.g., CuO, BN, or Cu$_2$N \cite{kahle2012,lothNP2010,tsukahara2009,loth2012}, the coherence times remain in the order of hundreds of ps. The separating layers cause the formation of an effective band gap in the substrate which results in an increased coherence time.

To take the concept of band gap introduction a step further, it was suggested to use superconducting substrate in which a perfect band gap for electrons is obtained \cite{Heinrich2013}. Magnetic defects adsorbed directly onto the superconducting surface would, however, generate undesired in-gap resonances, often referred to as Shiba states \cite{shiba1968,yasdani1997,ji2008,franke2011}. Such resonances can be avoided by using paramagnetic organic molecules, in which the ligands tend to separate the central magnetic ion from the conducting environment \cite{chen2008,Heinrich2013}. Accordingly, adsorbed, e.g., M-octaethylporphyrin-chloride (M-OEP-Cl) and M-phtalocyanine molecules, where M denotes a transition metal element (Mn, Fe, Co, Ni, Cu), onto a Pb(111) surface at low temperatures are suitable candidates. In addition, the organic skeleton typically generates an anisotropy field which acts on the magnetic moment, thereby creating a non-degenerate ionic spin structure which can be resolved with inelastic electron tunneling spectroscopy (IETS).

\begin{figure}[t]
\begin{center}
\includegraphics[width=0.75\columnwidth]{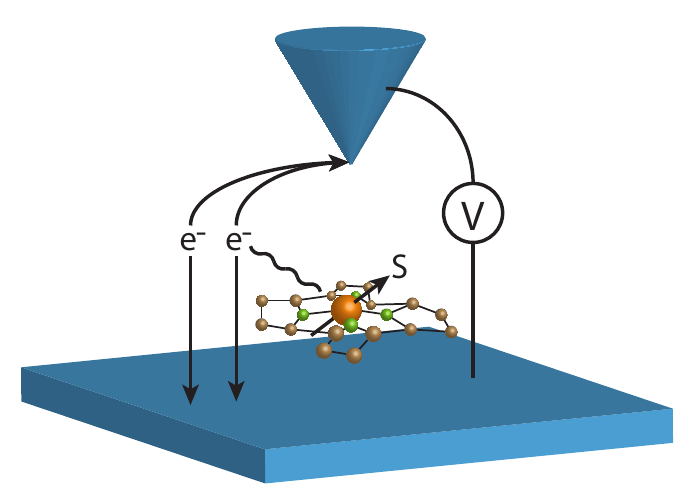}
\caption{Cartoon of the system set-up with superconducting scanning tunneling tip and substrate. Electrons $e^-$ tunnel in presence of the localized magnetic moment $\bfS$ with which they may or may not interact via exchange.}
\label{fig1}
\end{center}
\end{figure}

In this Letter, we theoretically investigate the properties and signatures of the electron tunneling conductance in a superconducting scanning tunneling microscopy/spectroscopy (STM/STS) set-up, see Fig. \ref{fig1}. Modeling the tunneling processes between the tip and substrate in presence of a localized magnetic moment using spin exchange interactions, we provide a transparent formalism for the analysis of the conductance spectra. In excellent agreement with experiment, we obtain inelastic spin transitions only outside of the superconducting tunneling gap at low temperatures. By varying anisotropies and magnetic fields, we investigate the properties of the conductance peaks in terms of the resulting excitation spectrum of the localized magnetic moment.

The superconducting gap formed in the substrate at low temperatures serves as an ideal structure for protecting excited spin states. This was demonstrated by superconducting STM/STS measurements \cite{Heinrich2013}, using Pb substrate and Pb covered scanning tip at 1.2 K. The experiments revealed that the spin excitations were accessible only for voltage biases $V$ larger than the sum of the superconducting gaps of the tip, $\Delta_{\rm tip}$, and the substrate, $\Delta_{\rm sub}$ \cite{Heinrich2013}. Hence, although spin excitations may be energetically within the gap, there are no single electron tunneling events that can facilitate transitions between the spin states of the magnetic ion for $|eV|<\Delta_{\rm tip}+\Delta_{\rm sub}$. For larger voltage biases the signatures of inelastic scattering is obtained in the (differential) conductance at $|eV|=\Delta_{\rm tip}+\Delta_{\rm sub}+\Delta_{mn}$, where $\Delta_{mn}=E_m-E_n$ is the energy difference between spin states of the magnetic ion. However, as the tunneling current only can exchange an angular momentum of $\Delta m_z=0,\pm1$, through spin preserving and spin flip scattering, with a sufficiently large probability, inelastic transitions can only occur between states which preserve the total angular momentum of the localized magnetic moment and the tunneling electrons.

Typically, signatures of inelastic scattering appear in the conductance spectra as an imprint of the underlying electronic structure. Thus, in measurements with normal metals without specific features in the density of electron states (DOS), the inelastic signatures generate simple steps in the conductance spectrum \cite{hirjibehedin2006,hirjibehedin2007,balashov2009,khajetoorians2010}, while tunneling through, e.g., double barrier structures provide inelastic side peak copies of the central electronic peak \cite{koch2005,okabayashi2007,galperin2007}. In the case of superconducting tip and substrate one would expect the gapped electronic structure with sharp coherence peaks to be replicated by the inelastic transitions \cite{balatsky2003}. Indeed, the recorded conductance spectra in Ref. \cite{Heinrich2013} very prominently show this feature.

For a phenomenological microscopical model of the set-up we propose the Hamiltonian
\begin{equation}
\Hamil=
	\Hamil_{\rm tip}
	+\Hamil_{\rm sub}
	+\Hamil_T
	+\Hamil_S,
\end{equation}
where $\Hamil_{\rm tip}=\sum_{\bfp\sigma}\dote{\bfp}\cdagger{\bfp}\cc{\bfp}+\sum_\bfp(\Delta_{\rm tip}\csdagger{\bfp\up}\csdagger{-\bfp\down}+H.c.)$ and $\Hamil_{\rm sub}=\sum_{\bfk\sigma}\dote{\bfk}\cdagger{\bfk}\cc{\bfk}+\sum_\bfk(\Delta_{\rm sub}\csdagger{\bfk\up}\csdagger{-\bfk\down}+H.c.)$ describe the superconducting tip and substrate, respectively. The operator $\cdagger{\bfq}$ ($\cc{\bfq}$) creates (annihilates) an electron at the energy $\leade{\bfq}$, momentum $\bfq$, and spin $\sigma=\uparrow\downarrow$, and we designate $\bfq=\bfp$ ($\bfk$) for states in the tip (substrate). The pairing potential for the tip (substrate) is denoted by $\Delta_{\rm tip(sub)}$. The single electron tunneling is described by $\Hamil_T=\sum_{\bfk\bfp\sigma\sigma'}\cdagger{\bfp}(\delta_{\sigma\sigma'}T_0+T_1\bfsigma_{\sigma\sigma'}\cdot\bfS)\cs{\bfk\sigma'}+H.c.$. While the first contribution in $\Hamil_T$ corresponds to single electron tunneling processes unaffected by the local magnetic moment $\bfS$, the second contribution provides electronic tunneling processes in which tunneling electrons are subject to magnetic exchange interaction with $\bfS$. The local magnetic moment embedded in the anisotropic environment of the organic molecule is modeled by the well known expression
\begin{equation}
\Hamil_S=
	-g\mu_B\bfB\cdot\bfS
	+DS_z^2
	+\frac{E}{2}(S_+^2+S_-^2),
\label{eq-Hs}
\end{equation}
where $g$ and $\mu_B$ is the gyromagnetic ratio and Bohr magneton, respectively, whereas $\bfB$ is the external magnetic field. The parameters $D$ and $E$ are the uniaxial and transverse anisotropy fields, respectively, exerted by the organic molecule and acting on the local magnetic moment. In general, the spectrum of this model can be decomposed into the eigensystem $\{E_\alpha,|\alpha\rangle\}$ of the $2S+1$ eigenenergies and eigenstates.

\begin{figure}[t]
\begin{center}
\includegraphics[width=0.99\columnwidth]{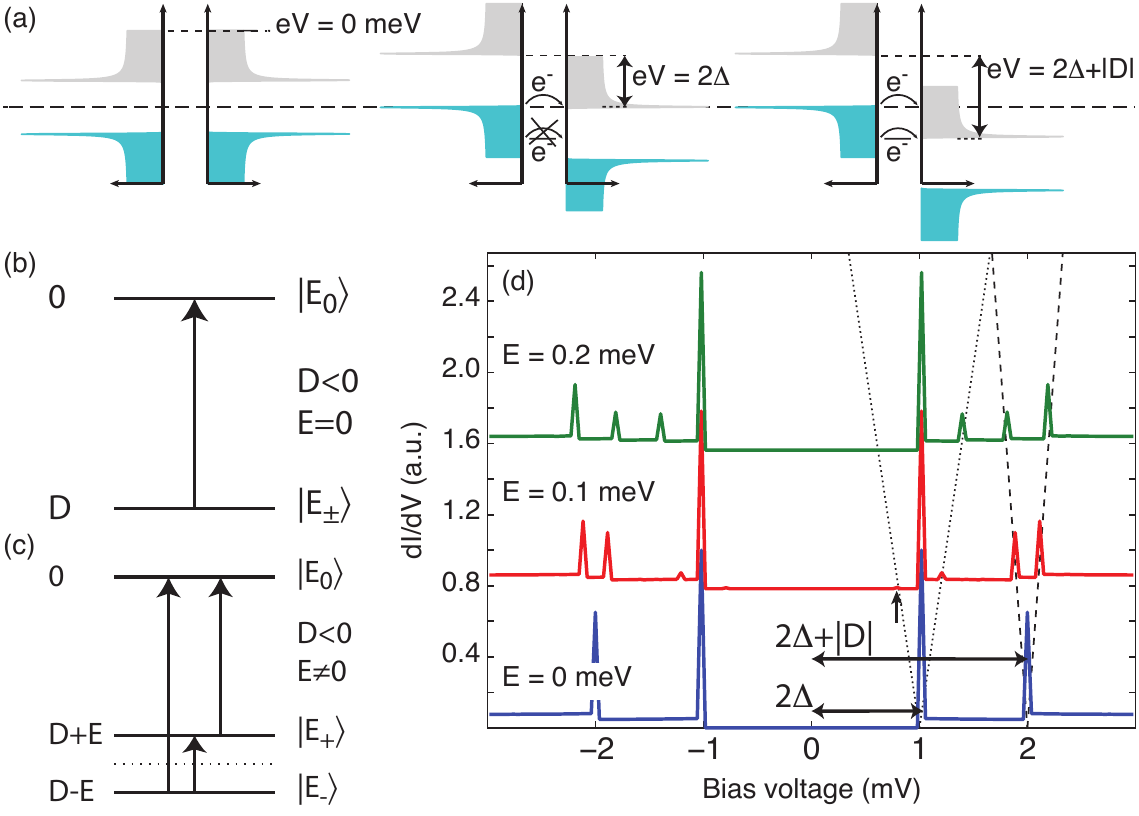}
\end{center}
\caption{(a) Schematic picture of the tunneling processes for the different voltage biases. (b), (c) Possible transitions in the spin system without (b) and with (c) transverse anisotropy field $E$. (d) Calculated conductance for a spin $S=1$ at $T=1.2$ K, for $D=-1$ meV and varying $E$. Here, also $\Delta_{\rm tip/sub}=0.5$ meV, $T_1=0.3T_0$, and $T=1.2$ K. The dashed (dotted) lines highlight the voltage biases for the transitions $|E_\pm\rangle\rightarrow|E_0\rangle$ ($|E_\pm\rangle\rightarrow|E_\mp\rangle$).
}
\label{fig-S1}
\end{figure}

The electron tunneling current flowing between the tip and substrate can be calculated as the rate of change of the electronic occupation in the tip, i.e., $I=-e\partial_t\sum_{\bfp\sigma}\langle\cdagger{\bfp}\cc{\bfp}\rangle$. Here and henceforth we shall omit the Josephson current since it does not pertain to the measurements of interest in the present study. For a heuristic derivation, neglecting scattering between states within the same lead but with different momentum and spin, e.g., processes like $\ket{\bfk\sigma}\bra{\bfk'\sigma'}$, the electron tunneling current can be written as
\begin{eqnarray}
I&=&
	-\frac{2e}{\hbar}\im\sum_{\bfk\bfp}\sum_{\sigma\sigma'}
	(-i)\int_{-\infty}^tdt'
\nonumber\\&&\times
		\av{\com{(\cdagger{\bfp}\hat{T}_{\sigma\sigma'}\cs{\bfk\sigma'})(t)}{(\csdagger{\bfk\sigma'}\hat{T}_{\sigma'\sigma}\cs{\bfp\sigma})(t')}}	
\end{eqnarray}
to second order tunneling processes, where $\hat{T}_{\sigma\sigma'}(t)=T_0\delta_{\sigma\sigma'}+T_1\bfS(t)\cdot\bfsigma_{\sigma\sigma'}$ is the tunneling operator. Using standard, e.g., non-equilibrium Green function methods, it is straightforward to derive the single electron current for the stationary conditions defined in the system. By resolving the spin-operator into its components, one shows in analogy to previous studies in the context of normal metals \cite{fransson2009,rossier2009,persson2009,lorente2009,fransson2010}, that the tunneling current can be be partitioned into three contributions $I=\sum_{n=0,1,2}I_n$. The first contribution $I_0\propto T_0^2$ provides a finite background current which does not couple to the local magnetic moment. Under the stationary conditions assumed for a constant voltage bias, this contribution can be written as
\begin{equation}
I_0=
	-\frac{4e}{\hbar}T_0^2\im\sum_{\bfk\bfp}
	\int
		\frac{
			g_\bfp^<(\omega)g_\bfk^>(\omega')
			-
			g_\bfp^>(\omega)g_\bfk^<(\omega')
		}
		{\omega-\omega'+i\delta}
	\frac{d\omega}{2\pi}\frac{d\omega'}{2\pi}
	,
\label{eq-I0}
\end{equation}
where $g_{\bfq}^{</>}(\omega)$ is the lesser/greater electron Green function for the superconducting tip and substrate. Assuming that the superconducting electronic structure of both the tip and substrate are unperturbed by the presence of the localized spin moment, we write the lesser/greater GFs as $g_\bfq^{</>}(\omega)=(\pm i)2\pi f_{\rm tip/sub}(\pm\omega)[|u_\bfq|^2\delta(\omega-E_\bfq)+|v_\bfq|^2\delta(\omega+E_\bfq)]$, with coherence factors $|u_\bfq|^2=(1+\dote{\bfq}/E_\bfq)/2$ and $|v_\bfq|^2=(1-\dote{\bfq}/E_\bfq)/2$, whereas $E_\bfq=\sqrt{\dote{\bfq}^2+|\Delta_{\rm tip/sub}|^2}$ denotes the quasi-particle energy. Here, also $f_{\rm tip/sub}(\omega)=f(\omega-\mu_{\rm tip/sub})$ is the Fermi function at the chemical potential of the tip/substrate, such that $\mu_{\rm tip}=\mu_{\rm sub}+eV$. Substituting into Eq. (\ref{eq-I0}), we obtain
\begin{align}
I_0=&
	\frac{4e\pi}{\hbar}T_0^2\sum_{\bfk\bfp}
	\biggl(
		[
			f(E_\bfp)-f(E_\bfk)
		]
		\Big(
			|u_\bfp|^2|u_\bfk|^2\delta(E_\bfp-E_\bfk-eV)
\nonumber\\&
			-|v_\bfp|^2|v_\bfk|^2\delta(E_\bfp-E_\bfk+eV)
		\Bigr)
		+
		[
			1-f_{\rm tip}(E_\bfp)-f(E_\bfk)
		]
\nonumber\\&\times
		\Big(
			|v_\bfp|^2|u_\bfk|^2\delta(E_\bfp+E_\bfk+eV)
			-|u_\bfp|^2|v_\bfk|^2\delta(E_\bfp+E_\bfk-eV)
		\Bigr)
	\biggr)
	.
\end{align}
Following the procedure outlined in, e.g., Ref. \onlinecite{mahan1990}, we can finally write this contribution to the tunneling current as
\begin{align}
I_0=&
	\frac{4e\pi}{\hbar}T_0^2
	\int
		\frac{nE\theta(E-|\Delta_t|)}{\sqrt{E^2-|\Delta_t|^2}}
		\frac{NE'\theta(E'-|\Delta_s|)}{\sqrt{E'^2-|\Delta_s|^2}}
		\biggl(
			[
				f(E)-f(E')
			]
\nonumber\\&\times
			\Big(
				\delta(E-E'-eV)
				-\delta(E-E'+eV)
			\Bigr)
			-
			[
				1-f(E)-f(E')
			]
\nonumber\\&\times
			\Big(
				\delta(E+E'+eV)
				+\delta(E+E'-eV)
			\Bigr)
		\biggr)
	dEdE',
\end{align}
where $n$ ($N$) is the density of electron states (DOS) in the tip (substrate). The corresponding conductance $dI_0/dV$ displays coherence peaks around $|eV|=\Delta_{\rm tip}+\Delta_{\rm sub}$, which are associated with the coherence resonances in the tip and substrate DOS, c.f. Fig. \ref{fig-S1} (a).

While the second contribution $I_1\propto T_0T_1$ does couple to the magnetic moment, it does not contribute to the electron current and is therefore omitted \cite{berggren2014}. Our main concern is with the third contribution $I_2\propto T_1^2$, derived under the same conditions and assumptions as $I_0$, since it contains information about the local spin fluctuations $\bfchi\sim\langle\bfS\bfS\rangle$. This is intelligible from the expression
\begin{widetext}
\begin{align}
I_2(V)=&
	i\frac{4\pi e}{\hbar}T^2_1
	{\rm sp}
	\int
		\frac{nE\theta(E-|\Delta_t|)}{\sqrt{E^2-|\Delta_{t}|^2}}
		\frac{NE'\theta(E'-|\Delta_s|)}{\sqrt{E'^2-|\Delta_{s}|^2}}
	\bfsigma\cdot
	\biggl\{
		\Bigl(
			f(E)f(E')\bfchi^>(\omega)
			-f(-E)f(-E')\bfchi^<(\omega)		
		\Bigr)
		\delta(E+E'-\omega+eV)
\nonumber\\&
		+
		\Bigl(
			f(-E)f(-E')\bfchi^>(\omega)
			-f(E)f(E')\bfchi^<(\omega)
		\Bigr)
		\delta(E+E'+\omega-eV)
		+
		\Bigl(
			f(E)f(-E')\bfchi^>(\omega)
			-f(-E)f(E')\bfchi^<(\omega)
		\Bigr)\delta(E-E'-\omega+eV)
\nonumber\\&
		+
		\Bigl(
			f(-E)f(E')\bfchi^>(\omega)
			-f(E)f(-E')\bfchi^<(\omega)
		\bigr)
		\delta(E-E'+\omega-eV)
	\biggr\}
	\cdot\bfsigma
	dEdE'd\omega,
\label{eq-I2}
\end{align}
\end{widetext}
where ${\rm sp}$ denotes the trace over the electronic spin degrees of freedom. Here, we have defined the spin-spin correlation functions $\bfchi^>(\omega)=(-i)\int\langle\bfS(t)\bfS(t')\rangle e^{-i\omega(t-t')}dt'$ and $\bfchi^<(\omega)=(-i)\int\langle\bfS(t')\bfS(t)\rangle e^{-i\omega(t-t')}dt'$. As we discuss below, the spin-spin correlation functions provide the spectrum of the spin transitions at the local moment, weighted by the populations of the states involved in the transitions, that is, $\bfchi^{</>}(\omega)=(-i)2\pi\sum_{mn}{\bf P}_{mn}\delta(\omega\mp\Delta_{mn})$, c.f. Eq. (\ref{eq-chii}).

The expression in Eq. (\ref{eq-I2}) constitutes the contribution to the tunneling current, and accordingly to the (differential) conductance $dI_2/dV$, that carries the signatures from the spin fluctuations. First, at zero temperature the Fermi functions $f(x)=0$ and $f(-x)=1$, which implies that only the first and second contributions in Eq. (\ref{eq-I2}) are finite. This leads to a finite $I_2$ for voltage biases satisfying $|eV-\omega|\geq\Delta_{\rm tip}+\Delta_{\rm sub}$, provided that $\bfchi^{</>}\neq0$. The conductance $dI_2/dV$ repeats the characteristic coherence peaks for voltage biases $|eV-\Delta_{mn}|\sim\Delta_{\rm tip}+\Delta_{\rm sub}$, as additional channels for conduction open up when the energetic tunneling electrons assist inelastic spin excitation transitions in the local moment. The energy fed into the tunneling current by the voltage bias can be absorbed by the local moment whenever it matches the transition energy $\Delta_{mn}\ (+\Delta_{\rm tip}+\Delta_{\rm sub})$. This process is reflected in the conductance as a new resonance peak with the same shape as the coherence peaks. The physical tunneling processes for different voltage biases are schematically depicted in Fig. \ref{fig-S1} (a), where the left panel refers to equilibrium conditions, whereas the middle panel illustrate the case $\Delta_{\rm tip}+\Delta_{\rm sub}\leq|eV|\leq\Delta_{\rm tip}+\Delta_{\rm sub}+\Delta_{mn}$ where tunneling between the tip and substrate which is not coupled to the local spin moment is possible, and $|eV|\geq\Delta_{\rm tip}+\Delta_{\rm sub}+\Delta_{mn}$ in the right panel.


Second, for finite temperatures all contributions in Eq. (\ref{eq-I2}) are non-vanishing. The additional, third and fourth, contributions describe de-excitation, or emission, processes of the local magnetic moment which open conductance channels. The energy emitted by the local moment in the de-excitation process may be absorbed by the tunneling electrons and open a new channel for conduction whenever the energy $|eV-\Delta_{mn}|=\Delta_{\rm tip}+\Delta_{\rm sub}$, and since the de-excitation energy $\Delta_{mn}<0$ there emerge new conductance resonances within the voltage bias range $\pm(\Delta_{\rm tip}+\Delta_{\rm sub})/e$. This will only occur, however, for elevated temperatures, $k_BT\gtrsim|\Delta_{mn}|$, that can sustain thermal excitations and de-excitations of the local moment.

Spin fluctuations in the sample play a crucial role in the present study. Using the model for the local spin moment introduced in Eq. (\ref{eq-Hs}), we expand the spin-spin correlation functions in terms of the eigensystem $\{E_\alpha,|\alpha\rangle\}$. Before proceeding, however, we notice that
\begin{align}
\label{eq-chisum}
\sum_{\sigma\sigma'}\bfsigma_{\sigma\sigma'}\cdot\bfchi^{</>}\cdot\boldsymbol{\sigma}_{\sigma'\sigma}&
	=2\chi^{</>}_z+\chi^{</>}_{-+}+\chi^{</>}_{+-},
\\
\label{eq-chii}
\chi_i^{</>}(\omega)=&
	(-i)2\pi\sum_{\alpha\beta}P_{\alpha\beta}^i\delta(\omega\mp\Delta_{\beta\alpha}),\ i=z,\pm\mp,
\end{align}
with the population factors $P_{\alpha\beta}^z=\langle\alpha|S_z|\beta\rangle(1-P_\beta)\langle\beta|S_z|\alpha\rangle P_\alpha$ and $P_{\alpha\beta}^{\mp\pm}=\langle\alpha|S_\mp|\beta\rangle(1-P_\beta)\langle\beta|S_\pm|\alpha\rangle P_\alpha$, where $P_\alpha$ is the population of the state $|\alpha\rangle$. The occupation numbers $P_\alpha$ can be provided through, e.g., the Fermi-Dirac or Gibbs distribution.

It is instructive to study an example with $S=1$, which is characterized by the eigensystem $\{E_0=0,|E_0\rangle; E_\pm=D,|E_{\pm1}\rangle\}$, where $|E_{m_z}\rangle=|m_z=0,\pm1\rangle$ for $E=0$. The spin-preserving expectation values $\langle\alpha|S^z|\beta\rangle=0$, $\alpha\neq\beta$, which is clear since the states $|m_z=0\rangle$, $|m_z=\pm1\rangle$, are decoupled. Hence, the only transitions that contribute to the conductance are the spin changing transitions since expectation values of the type $\langle m_z=0|S^\pm|m_z=\mp1\rangle\neq0$, as schematically depicted in Fig. \ref{fig-S1} (b). These transitions have to be accompanied by tunneling electrons that undergo spin flips, in order for the system to preserve its total angular momentum, and provide the energy required for the transition. Conductance peaks will therefore only emerge at $|eV|=\Delta_{\rm tip}+\Delta_{\rm sub}+|D|$, which is illustrated in the calculated $dI/dV$ shown in the bottom of Fig. \ref{fig-S1} (d). The plot clearly shows that the coherence peaks at $eV\sim\pm1$ mV are replicated by the inelastic scattering signal at $eV\sim\pm2$ mV.

For $E\neq0$, the eigensystem is modified by $E_\pm=D\pm E$, $|E_{\pm1}\rangle\equiv[|m_z=-1\rangle\pm|m_z=1\rangle]/\sqrt{2}$, which breaks the degeneracy of the states $|E_{\pm1}\rangle$ and splits up the energy of $|E_{+1}\rangle$ and $|E_{-1}\rangle$ by $2E$. The spin changing transitions, e.g., $\langle E_0|S_+|E_-\rangle$ and $\langle E_0|S_-|E_+\rangle$ therefore occur at different energies, see Fig. \ref{fig-S1} (c), and we expect signatures in the conductance at the voltage biases $|eV|=\Delta_{\rm tip}+\Delta_{\rm sub}+D\pm E$, which is readily seen in Fig. \ref{fig-S1} (d). In addition, since the Fock states $|m_z=\pm1\rangle$  are coupled, the tunneling current also facilitates spin-preserving transitions between the states $|E_{+1}\rangle$ and $|E_{-1}\rangle$. Inelastic signatures of these transitions are expected to appear on both sides of the main coherence peaks at $|eV|=\Delta_{\rm tip}+\Delta_{\rm sub}\pm E$, see Fig. \ref{fig-S1} (d). The middle plot ($E=0.1$ meV) also displays in-gap absorption transitions, indicated by the arrow, since $k_BT\sim0.1$ meV is sufficient energy for the states $|E_{\pm1}\rangle$ to be thermally excited and, hence, accessible for tunneling assisted transitions.

The apparent difference in amplitude between the transitions $|E_{\pm1}\rangle\rightarrow|E_0\rangle$ and $|E_{\pm1}\rangle\rightarrow|E_{\mp1}\rangle$, which is legible from Fig. \ref{fig-S1} (d), can be understood in terms of the population factors $P_{\alpha\beta}$. For $D<0$ and small $E\neq0$, the populations $P_\pm$ of the states $|E_{\pm1}\rangle$ are both close to 1, such that, e.g., $(1-P_+)P_-$ becomes small. The population $P_0$ for the state $|E_0\rangle$ is, on the other hand, small which leads to relatively large products $(1-P_0)P_\pm$.

Next, we connect to recent experiment observations by turning our attention to the spin $S=5/2$ system \cite{Heinrich2013}. For $E=0$, the eigensystem consists of the doubly degenerate states $|m_z=\pm m/2\rangle$, $m=1,3,5$, at energies $E_{\pm m/2}=Dm^2/4$, and with a positive (negative) uniaxial anisotropy, $D>0$ ($D<0$), the system acquires a minimal (maximal) spin state $|\pm1/2\rangle$ ($|\pm5/2\rangle$).

\begin{figure}[t]
\begin{center}
\includegraphics[width=0.99\columnwidth]{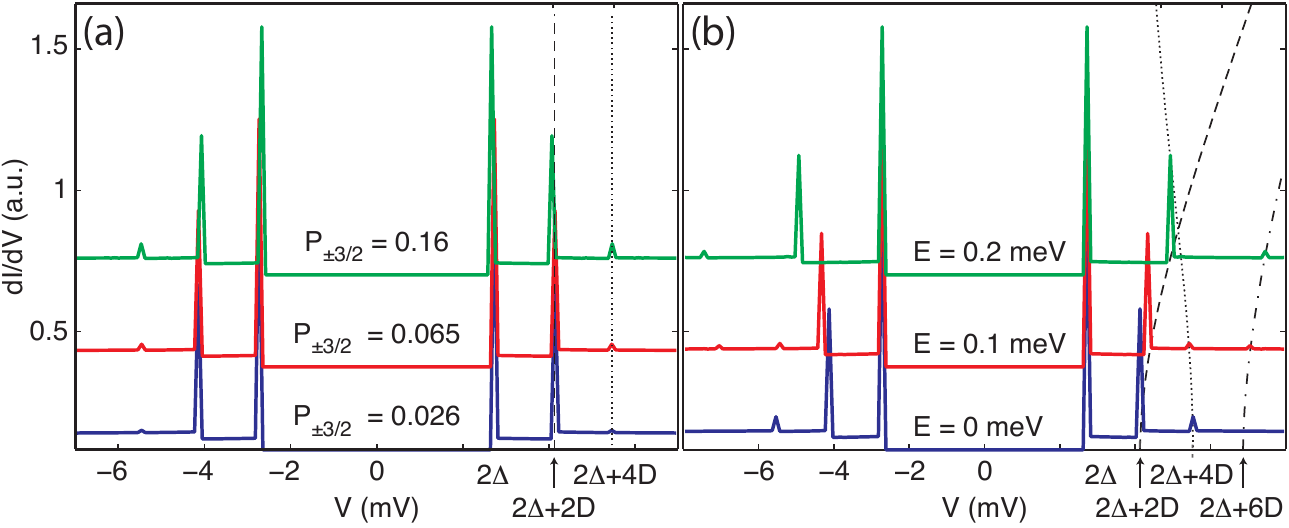}
\caption{Calculated conductances for a spin $S=5/2$ system under varying (a) population of the states $|m_z=\pm3/2\rangle$ for $E=0$ and (b) transverse anisotropy $E$, $\mu=-1.4$ meV. Other parameters are $D=0.7$ meV, $\Delta_{\rm tip/sub}=1.35$ meV, $T=1.2$ K \cite{Heinrich2013}, and $T_1=0.3T_0$. Notice the different horizontal scales in panels (a) and (b).}
\label{fig-S2}
\end{center}
\end{figure}

In Fig. \ref{fig-S2} (a) we plot the calculated conductance for varied populations of the states $|m_z=\pm3/2\rangle$ in absence of transverse anisotropy, $E=0$. Here, we change the population of these states by rigidly shifting the spin spectrum relative to the chemical potential of the system in equilibrium. Although such effects of increasing the populations of excited states can be accounted for in more sophisticated ways \cite{novaes2010}, we apply this simplistic methodology which is sufficient to illustrate the physics of pumping, in very good agreement with experiment. Here, we assume that the pairing potentials of the tip and substrate are equal, $\Delta_{\rm tip/sub}=\Delta\sim1.35$ meV, neglecting possible superconducting phase difference, and positive uniaxial anisotropy $D=0.7$ meV. Analogous to the previous case, the conductances display strong coherence peaks at $eV=\pm2\Delta$, which are perfectly replicated at the voltage biases $|eV|=2\Delta+2D$ for the inelastic spin transition $|m_z=\pm1/2\rangle\rightarrow|m_z=\pm3/2\rangle$.

We, furthermore, notice the conductance peak emerging at voltage biases $|eV|=2\Delta+4D$ for increasing population of the first excited states $|m_z=\pm3/2\rangle$. The conductance peak is a signature of the inelastic transition $|m_z=\pm3/2\rangle\rightarrow|m_z=\pm5/2\rangle$ and its characteristics can be quantified by using the expressions in Eqs. (\ref{eq-chisum}) and (\ref{eq-chii}). As the matrix elements for raising and lowering between the states $|m_z=\pm3/2\rangle$ and $|m_z=\pm5/2\rangle$ are always finite in the present set-up, the emergence of the conductance peak strongly depends on the population of these states. When in the ground state, both $|m_z=\pm3/2\rangle$ and $|m_z=\pm5/2\rangle$ are unpopulated which leads to vanishing factors $P_{\pm\frac{3}{2}\pm\frac{5}{2}}$. This scenario remains valid for small charge currents through the system, as well. For increasing charge currents, however, density accumulates in the states $|m_z=\pm3/2\rangle$ as they are excited with a faster rate than their corresponding decoherence times. Accordingly, upon populating those states, the factors $P_{\pm\frac{3}{2}\pm\frac{5}{2}}$ become finite which leads to additional conduction channels open as the transitions $|m_z=\pm3/2\rangle\rightarrow|m_z=\pm5/2\rangle$ become available.

In this fashion we reproduce the effect of pumping which is obtained in the experiment by decreasing the distance between the scanning tip and the sample. Decreasing the tip-sample distance, however, also appears to have a strong influence on the uniaxial anisotropy since a shift in the excitations was observed \cite{Heinrich2013}. As the microscopic details of this feature are unknown and its origin is beyond the scope of the present Letter, we have omitted this shift in our calculated spectra.

Next, we observe in Fig. \ref{fig-S2} (b), that a finite transverse anisotropy $E$ generates a modified spectrum for the spin system. Although the states remain doubly degenerate, they become linear combinations of the kind $|E_{\pm m}\rangle=\sum_{n=1,3,5}\alpha_{\pm n/2}^{(m)}|m_z=\pm n/2\rangle$. As an effect, the ground state comprise not only the low spin Fock states but also the higher spin Fock states. Hence, while the ground state of the spin is populated there is finite probability that, e.g., the raising operator $S_+$ generates transitions to spin Fock states that are not permitted in absence of the transverse anisotropy. This can be seen in Fig. \ref{fig-S2} (b) along the dash-dotted trace, which marks a resonance that appears due to transitions between the ground state and the highest excitation which are mainly weighted on $|m_z=\pm1/2\rangle$ and $|m_z=\pm5/2\rangle$, respectively. Despite seemingly violating spin conservation, these transitions  are permitted for $E\neq0$, since both the ground and highest excited states are linear combinations of the Fock states $|m_z=\pm1/2\rangle$, $|m_z=\pm3/2\rangle$, and $|m_z\pm5/2\rangle$. Consequently, there is density distributed among the Fock states that enable transitions with $\Delta m_z=\pm1$.

\begin{figure}[t]
\begin{center}
\includegraphics[width=0.75\columnwidth]{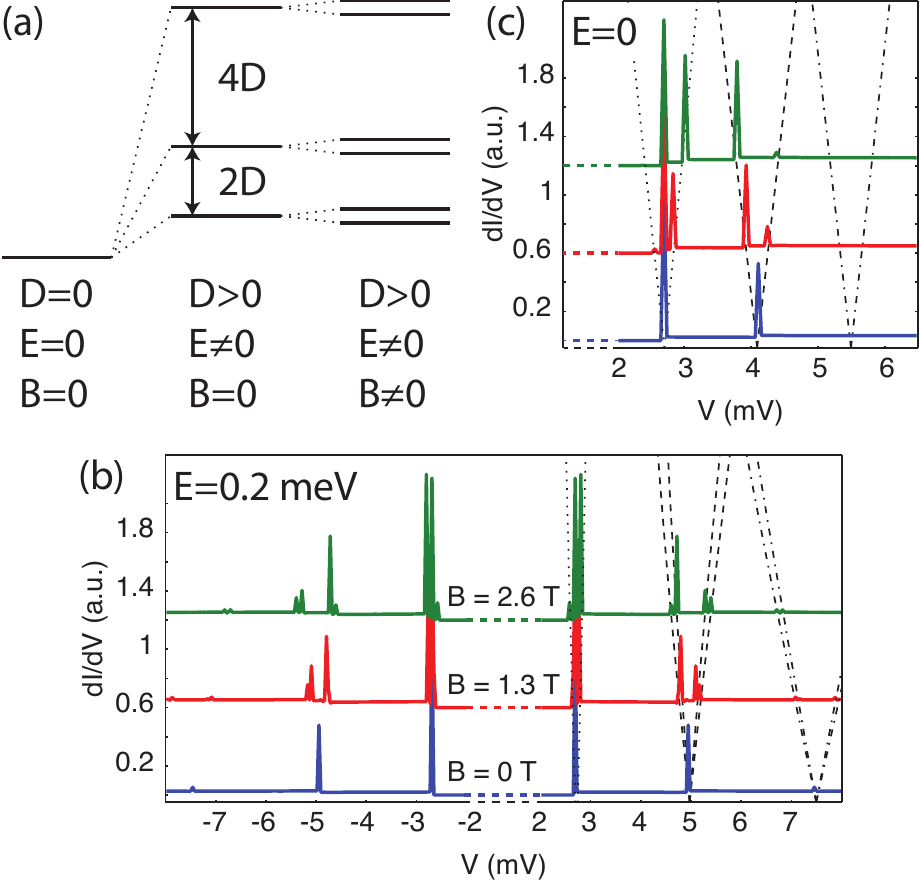}
\caption{(a) Spectrum of a spin $S=5/2$ system in the atomic limit subject to different conditions, parametrized by the anisotropies $D, E$ and external magnetic field ${\bf B}=B\hat{\bf z}$. (b) Conductance spectra for increasing $B$ and $E=0.2$ meV. (c) Same as in panel (b) for $E=0$ and for positive voltage biases only. Other parameters are as in Fig. \ref{fig-S1}.}
\label{fig-S3}
\end{center}
\end{figure}

Finally, we consider the spin $S=5/2$ under moderate magnetic fields in order to elucidate and reveal further information about the excitation spectrum of the local magnetic moment. As we have seen in the above, the uniaxial and transverse anisotropies, $D$ and $E$, are insufficient to break up the two-fold degeneracies in the spectrum and an external magnetic field has to be supplied to achieve such separation, see Fig. \ref{fig-S3} (a).

The plots shown in Fig. \ref{fig-S3} (b) display the expected conductance spectra as the remaining degeneracies are broken due to the external magnetic field. First, we notice the emergence of additional peaks surrounding the superconducting coherence peaks, with increasing magnetic field, due to inelastic emission and absorption scattering between the Zeeman split ground state(s). Second, the first excited peak (emerging around $V\sim\pm5$ mV in the lower panel) is split into four since there are four possible transitions allowed for finite $E$ between the Zeeman split ground and first excited states. Finally, transitions between the Zeeman split ground and second excited states are visible in the spectra (emerging around $V\sim\pm7.5$ mV in the lower panel).

We notice here that absence of the transverse anisotropy field $E$, removes the coupling between spin projections, such that only transitions like, e.g., $\langle m_z=3/2|S_+|m_z=1/2\rangle$ and $\langle m_z=-3/2|S_+|m_z=-1/2\rangle$, are possible. Under the magnetic field, these occur at different energies due to the different Zeeman split of the ground and first excited states. Hence, only two peaks appear in the conductance spectra (at around $V\sim\pm4$ mV), which can be seen in Fig. \ref{fig-S3} (c).

Providing an external magnetic field adds a complication to the measurements, since the superconductivity in the both substrate and tip becomes quenched under too strong fields. This problem can, however, be overcome by changing to a tip/substrate material that is less sensitive to magnetic fields, e.g., NbTi, Nb$_3$(Sn,Ge,Al), and MgB$_2$ \cite{larbalestier2001,buzea2001,gurevich2011}, which are known to maintain their superconducting phase for fields as strong as 10-30 T. Our predictions made for fields up to a few T are therefore safely within the realms of feasibility.

We conclude this Letter by noticing that while a simple exchange coupling between the tunneling electrons and the localized magnetic moment is sufficient for a sound description of the conductance spectra, Eq. (\ref{eq-I2}), taken of, e.g. Fe-OEP-Cl. Questions regarding spurious states \cite{shiba1968,yasdani1997,ji2008,franke2011} within the superconducting gap of the substrate and the tip cannot be addressed within our simple approach. Such states are often observed in conjunction with localized magnetic moments adsorbed directly onto a superconducting surface. In the experimental set-up this is achieved by isolating the transition metal atom from the superconducting electrodes through the ligand cage surrounding the magnetic moment. We can from our model, however, deduce that the basic physical mechanism is contained in the exchange interaction between the magnetic moment and the tunneling electrons. This conclusion can be drawn despite that we do not consider higher order influence from the spin fluctuations, since the tunneling current, Eq. (\ref{eq-I2}), provides a response to the spin excitations only outside the superconducting gaps of the tip and substrate, for low temperatures. We, furthermore, predict that more in-depth studies of the spin excitation spectrum can be undertaken by introducing external magnetic fields. However, as the superconducting properties may be quenched, care has to be taken concerning the choice of superconducting tip and substrate material, as well as concerning the strength of the magnetic field. Despite those possible complications, our predictions are within the realms of the state-of-the-art experimental technology.

\acknowledgments
We thank the Swedish Research Council for financial support.

\end{document}